\RequirePackage{lineno}
\documentclass[aps,prl,twocolumn,groupedaddress,showpacs,floatfix,letterpaper]{revtex4}
\usepackage{xcolor}
\usepackage{graphicx}
\usepackage{multirow}
\usepackage{amsfonts}
\usepackage{amsmath}
\usepackage{amssymb}
\usepackage{hyperref}
\usepackage{color}
\usepackage{multirow}
\usepackage{lineno}
\usepackage{siunitx}
\usepackage{accents}

\def\nue{\nu_e}
\def\numu{\nu_\mu}

\def\nutau{\nu_\tau}

\newcommand{\beq}{\begin{eqnarray}}
\newcommand{\eeq}{\end{eqnarray}}
\def\to{\rightarrow}

\def\attdiii{\accentset{\circ}{a}^{(3)}_{\tau\tau}} 
\def\supbaylimdimiiitattfowo{2 \times 10^{-26}}
\newcommand{\eVV}{\si{\electronvolt}}
\newcommand{\eVVmo}{\si{\electronvolt}\mathrm{^{-1}}}

\newcommand{\GeVV}{\si{\giga\electronvolt}}

\def\supbaylimdimiiitattfowo{2 \times 10^{-26}}
\def\testlimit{1 \times 10^{-26}}

\def\DMMW{4\times 10^{-28}}
\def\DM10yr{2\times 10^{-22}}
\def\DM1yr{2\times 10^{-23}}

\def\DMlocal{0.3} 
\def\DMFermi{10^{-13}} 
\def\DMNSI{8\times 10^{-9}} 
\def\DMV{3\times 10^{-33}} 
\def\DMA{3\times 10^{-13}} 
\def\AmpLimit{6\times 10^{-25}} 
 
\def\NuAdiabatic{1.6\times 10^{-29}}
\def\DMAdiabatic{4.0\times 10^{-11}}

\begin{document}

\newcommand{\makespace}{\vspace{3 mm}}
\newcommand{\DP}{\displaystyle}

\newcommand{\CA}[1]{\textcolor{blue}{{ [CA: \bf #1]}}}
\newcommand{\TK}[1]{\textcolor{red}{{ [TK: \bf #1]}}}
\newcommand{\KF}[1]{\textcolor{orange}{{ [KF: \bf #1]}}}

 \title{Ultra-light Dark Matter Limits from Astrophysical Neutrino Flavor}

\date{\today}

\smallskip
\smallskip
\author{Carlos~A.~Arg\"{u}elles$^{1}$,
Kareem Farrag$^{2}$, and
Teppei Katori$^{3}$\\
}

\smallskip
\smallskip
 \affiliation{
   $^1$Department of Physics \& Laboratory for Particle Physics and Cosmology, \\
Harvard University, Cambridge, MA 02138, USA \\
$^2$Department of Physics and Institute for Global Prominent Research, Chiba University, Chiba 263-8522, Japan\\
$^3$Department of Physics, King's College London, London WC2R 2LS, UK\\
 }

\begin{abstract}
Ultra-light dark matter is a class of dark matter models where the mass of the dark matter particle is very small and the dark matter behaves as a classical field pervading our galaxy.
If astrophysical neutrinos interact with ultra-light dark matter, these interactions would produce a matter potential in our galaxy which may cause anomalous flavor conversions.
Recently, IceCube high-energy starting event flavor measurements~\cite{IceCube:2020wum,IceCube:2020fpi} are used to set stringent limits on isotropic Lorentz violating fields under the Standard-Model Extension framework~\cite{IceCube:2017qyp,IceCube:2021tdn}. 
We apply the IceCube Lorentz violation limits to set limits on neutrino - ultra-light dark matter couplings. 
We assume the dark matter field undergoes fast oscillations in our galaxy, yielding neutrino interactions with dark matter that broaden and smear the observed flavor structure of astrophysical neutrinos at IceCube.
The constraints we obtain are an order of magnitude tighter than current and future terrestrial neutrino experimental limits.
The sensitivity of ultra-light dark matter can be further improved in the near future by new particle identification algorithms in IceCube and the emergence of next-generation neutrino telescopes.
\end{abstract}

\maketitle

{\it IceCube constraints on Lorentz violation} --- Quantum-gravity-motivated models allow for new spacetime structure in vacuum, such as Lorentz violation~\cite{Kostelecky:1988zi,Amelino-Camelia:1997ieq,Addazi:2021xuf}. 
So far, we have not observed such new spacetime structure~\cite{Kostelecky:2008ts}, with current limits suggesting that quantum gravitational effects are suppressed with the inverse of the Planck scale (1/$E_p\sim 10^{-19}$~GeV$^{-1}$) or larger powers. 

If such a structure affects neutrinos, then neutrino flavor information may contain imprints of such effects because any quantum mixings would alter the predicted behaviour of astrophysical neutrino mixing thanks to the tiny coupling between new physics and neutrinos. 

Such models describing this mixing with neutrinos are most generally formalised in the Standard-Model Extension (SME) framework, including more specific phenomena as Lorentz and CPT violation~\cite{Colladay:1996iz,Colladay:1998fq,Kostelecky:2003fs}. 
The framework contains a series of terms pertaining to the neutrino sector~\cite{Kostelecky:2003cr}, including higher order non-renormalizeable operators which dominate at higher neutrino energies~\cite{Kostelecky:2011gq}. 
Astrophysical neutrinos have the largest energy ($\geq 60$~TeV) and travel the longest distance from production to detection ($\geq$~Mpc), making them ideal candidates to probe these higher order operators.
By studying their flavor content, neutrinos are an ideal candidate for searches of new physics~\cite{Arguelles:2022tki}. 

Due to the small sample size of high-energy neutrino events, IceCube measures the flux-integrated flavor ratio of astrophysical neutrinos~\cite{IceCube:2020fpi,IceCube:2015rro,IceCube:2015gsk,IceCube:2018pgc}. 
Astrophysical neutrino flavor analyses currently have no charge sensitivity, and the flavor ratio includes both neutrinos and anti-neutrinos. 
The flavor ratio measured using the 7.5-yr high-energy starting event (HESE) sample~\cite{IceCube:2020wum,IceCube:2020fpi} was recently used to look for non-zero SME coefficients as a possible signature of Lorentz violation~\cite{IceCube:2021tdn}. 
Although this analysis did not find indications of Lorentz violation, it set the most stringent limits on several SME coefficients. 
These limits can be identified as those on any vector or axial-vector couplings with neutrinos.

The effective Hamiltonian of astrophysical neutrinos in the flavor basis can be written explicitly as
\begin{eqnarray}
\mathcal{H}_{\rm eff}
&=&
\frac{1}{2E}
\left(\begin{array}{ccc}
m^{2}_{ee} & m^{2}_{e\mu} & m^{2}_{\tau e} \\
m^{2*}_{e\mu} & m^{2}_{\mu\mu} & m^{2}_{\mu\tau}\\
m^{2*}_{\tau e} & m^{2*}_{\mu\tau} & m^{2}_{\tau\tau}
\end{array}\right)
+\left(\begin{array}{ccc}
\accentset{\circ}{a}^{(3)}_{ee} & \accentset{\circ}{a}^{(3)}_{e\mu} & \accentset{\circ}{a}^{(3)}_{\tau e} \\
{a}^{(3)}_{e\mu} & {a}^{(3)}_{\mu\mu} & {a}^{(3)}_{\mu\tau} \\
{a}^{(3)}_{\tau e} & {a}^{(3)}_{\mu\mu} & \accentset{\circ}{a}^{(3)}_{\tau\tau}
\end{array}\right) \nonumber \\
&=&V^\dagger \operatorname{diag}(\lambda_1,\lambda_2,\lambda_3)V.
\label{eq:hamiltonian}
\end{eqnarray}
Here, the first term is the neutrino mass matrix in the flavor basis~\cite{Esteban:2020cvm}, and the second is the dimension-three isotropic SME term~\cite{Kostelecky:2011gq}. 
Their sum makes an effective Hamiltonian, which can be diagonalized with a unitary matrix $V$. 
The flavor transition in the astrophysical case is phenomenologically equivalent to an incoherent sum due to a combination of the long propagation distance, energy resolution, and uncertainty in the production region, yielding that the mixing probability from the flavor state $\alpha$ to $\beta$ to be written as
\begin{eqnarray}
P_{\alpha\to\beta}(E)=\sum_i|V_{\alpha,i}|^2|V_{\beta,i}|^2~.\label{eq:osc}
\end{eqnarray}
Utilising 7.5 years of HESE flavor data, the IceCube analysis set its best limit on the dimension-three operator in the $\tau-\tau$ sector with respect to the null hypothesis at $\attdiii=\supbaylimdimiiitattfowo$~GeV and Bayes Factor $>10$.

In respect of this, we focus our investigation on the $\tau-\tau$ sector. 
This term is also the least constrained term from other experiments~\cite{Karmakar:2020yzn}, and it is interesting to set limits from astrophysical neutrino flavor data. 
In the future, with larger sample sizes and improved measurements, we will investigate other terms once IceCube either sets new limits or finds potential signals of such new physics phenomena as aforementioned.

{\it Methodology --- }In this Letter, we assume that the new physics interaction is flavor diagonal.
Since the neutrino mass matrix includes off-diagonal terms, one can expect modification of astrophysical neutrino flavor occurs through the unitary matrix $V$.
The effect of a nonzero SME coefficient is understood as a type of quantum Zeno effect~\cite{Harris:1980zi}.
In the presence of a nonzero $\tau-\tau$ element, this is equivalent to having $e-e$ and $\mu-\mu$ elements in the effective Hamiltonian.
Since the standard astrophysical neutrino production models include only electron neutrinos and muon neutrinos, this means nonzero $\tau-\tau$ element decouples tau neutrino flavor from electron neutrino and muon neutrino flavor content.
This is contradictory to current observation, where the best-fit disfavors both electron neutrino or muon neutrino dominant flavor at detection ~\cite{IceCube:2020fpi} as well as the recent claim of high-energy tau neutrino candidates by IceCube~\cite{IceCube:2024pov}.

Any potentials in the universe can act on astrophysical neutrino propagation and modify observed astrophysical neutrino flavor from the standard value.
Such potentials can be induced by new interactions between neutrinos and dark matter~\cite{Miranda:2013wla,Berlin:2016woy,Reynoso:2016hjr,Arguelles:2017atb,Choi:2019zxy,Chen:2023vkq}.
Ultra-light dark matter is a class of theoretically-motivated dark matter candidates where the mass of dark matter particles is very small, and it behaves like a classical field~\cite{Khlopov:1985fch,Marsh:2015xka}.
The vector and axial-vector couplings of dark matter and neutrinos, if they exist, would form a potential term that can be interpreted as the $\attdiii$ term to modify the neutrino flavor content.
They include couplings with vector mediators with scalar dark matter~\cite{deSalas:2016svi,Pandey:2018wvh,Smirnov:2019cae,Karmakar:2020yzn,Losada:2021bxx}, derivative couplings of scalar dark matter~\cite{Farzan:2018pnk,Gherghetta:2023myo,Cordero:2023hua}, contact neutrino - dark matter interactions~\cite{Salla:2022dxc}, couplings with vector dark matter~\cite{Brdar:2017kbt,Brzeminski:2022rkf,Alonso-Alvarez:2023tii,Lin:2023xyk}, and couplings with axion-like particles (ALP)~\cite{Huang:2018cwo,Reynoso:2022vrn}.

Notably, interactions between astrophysical neutrinos and ultra-light dark matter in our galaxy are of particular interest.
This is because these couplings to the dark matter would modify the neutrino flavor content locally, in regions where the dark matter does not exist, the standard Hamiltonian acting on neutrinos brings the flavor ratio close to the standard value given the short oscillation scale in comparison to astrophysical distances.
The propagation distance of neutrinos in the galaxy is always much longer than their oscillation length, therefore Eq.~\eqref{eq:osc} applies.
Note, we do not consider scalar coupling models, which can be interpreted as a modification of the neutrino mass term~\cite{Krnjaic:2017zlz,Liao:2018byh,Dev:2020kgz}.

Here, we assume astrophysical neutrino propagation through the extragalactic medium does not impact our results significantly.
This is due to the low background density of dark matter, resulting in approximately equally mixed neutrinos of all flavors in this scenario~\cite{Farzan:2008eg}.
Finally, we use $\rho\sim\DMlocal$~GeV/cm$^3$ as an averaged Milky Way dark matter density.
The precise density profile of dark matter may provide more stringent results, so our results presented here can be interpreted as conservative.
We further note that the HESE data sample is an all-sky, diffuse isotropic sample but it is still statistically limited.
IceCube maximized the statistical power by integrating over all directions to provide the best flavor result.
Through this process, it averages out the effect of the galactic density profile effect of the Milky Way galaxy.

{\it Ultra-light dark matter} --- A reasonable expectation for the missing mass of the universe~\cite{deSwart:2017heh}, or dark matter, is that they are undiscovered new particles.
Weakly interacting massive particles are a popular class of dark matter candidates, and experiments all over the world are designed to detect the interactions between galactic dark matter and low-energy threshold detector materials~\cite{NEWS-G:2017pxg,XENON:2018voc,PandaX-4T:2021bab,CRESST:2022lqw,LZ:2022ufs,DarkSide-50:2022qzh}.
On the other hand, ultra-light dark matter is a class of dark matter models, and its mass can be as small as $10^{-22}$~eV to account for all dark matter density in the galaxy, or of lower mass to make a part of the galactic dark matter density.
We highlight that if such ultra-light dark matter permeates galactic space and couple to astrophysical neutrinos, this can affect the IceCube astrophysical neutrino flavor data~\cite{Arguelles:2022tki}.

First, we consider a generic ultra-light dark matter $\phi$ with mass $m_\phi$, which interacts with neutrino with a vector mediator.
Since the IceCube limit is on the $\tau-\tau$ element, here we focus our attention tau neutrino - dark matter coupling.
We assume the time component dominates; this implies the dark matter field is isotropic in our rest frame.
Such a view can be justified, for example, if the field is at rest in the CMB frame where the cosmic microwave background (CMB) looks isotropic, and any spatial components are suppressed on the order of the galactic velocity, $\sim 10^{-3}$.
The potential arising from neutrino-dark matter interactions can be generalized as $V_{\tau\tau}=G'_{F,\tau\tau}\cdot \rho/m_\phi$~\cite{Karmakar:2020yzn} where $G'_{F,\tau\tau}$ is an effective Fermi constant-like coupling due to the new mediator.
Since we identify $\accentset{\circ}{a}^{(3)}_{\tau\tau}$ to be $V_{\tau\tau}$, we obtain,
\beq
G'_{F,\tau\tau}<\DMFermi~{\rm GeV}^{-2}\left(\frac{m_\phi}{10^{-20}\rm eV}\right).
\eeq
This limit can be identified as $g/\Lambda^2$ in~\cite{Farzan:2018pnk}.
The coupling is often normalized to the Fermi constant, in an analogous notation with non-standard interaction, $G'_F\equiv\epsilon_{\tau\tau}G_F$~\cite{deSalas:2016svi}.
Then, we obtain,
\beq
\epsilon_{\tau\tau}<\DMNSI\left(\frac{m_\phi}{10^{-20}\rm eV}\right).
\eeq

The classical field would oscillate with period $m_\phi$~\cite{Berlin:2016woy}, in this case, with unknown initial phase $\delta$,
\beq
\phi(t)=g_{\tau\tau}\frac{\sqrt{2\rho}}{m_\phi}\sin(m_\phi t+\delta).
\label{eq:vector}
\eeq
The limits can be obtained for the time-dependent potential via neutrino flavor data.
The longest time scale on terrestrial neutrino experiments is order a year, corresponding to $\sim\DM1yr$~eV, thus terrestrial neutrino experiments explore mainly $m_{\phi}\ge \DM1yr$~eV.
On the other hand, the time scale of astrophysical neutrinos is much longer.

Crossing the Milky Way galaxy takes over $5\times 10^4$ light years, the length scale corresponds to the wavelength of ultra-light dark matter $m_\phi\sim \DMMW$~eV.
Thus, astrophysical neutrinos with $m_\phi\ge \DMMW$~eV would experience many cycles of dark matter oscillation when they go through our galaxy.
Furthermore, the initial phase of the dark matter oscillation may be different for neutrinos from different directions.
So the phase of neutrino-dark matter coupling is then averaged out, and the average dark matter potential neutrinos experienced is mostly zero~\cite{Huang:2018cwo}.
However, even in this case, couplings with the background field would smear the flavor structure, and it will contribute to broadening the observed flavor ratio, a similar effect predicted in other systems~\cite {Hamaide:2022rwi}.

The key observation is that the light dark matter oscillation is fast compared with the time scale of neutrino propagation of the galaxy ($\sim 5\times 10^4$~years), and neutrinos interact with many cycles of oscillations; however, light dark matter oscillation is not fast enough to break the adiabatic neutrino flavor transition.
We use the solar neutrinos as a benchmark of the adiabaticity; then we obtain a conservative bound of the potential gradient, $\NuAdiabatic$~GeV/m, to maintain the adiabaticity of flavor conversion.

On the other hand, the IceCube sensitivity limit of neutrino - light dark matter coupling is given from their Lorentz violation search analysis, $\attdiii=\supbaylimdimiiitattfowo$~GeV.
We interpret this as an amplitude of the neutrino - dark matter coupling oscillation.
Then, we derive the adiabaticity condition is satisfied when the dark matter oscillation length is order $1$~km, or light dark matter mass $m_\phi<\DMAdiabatic$~eV.
Interestingly, the astrophysical neutrino flavor-sensitive region to search for neutrino-light dark matter coupling overlaps with terrestrial searches, and by combining them, one can look for such signals in a wide range of light dark matter masses.

\begin{figure}[t!]
 \begin{center}
\includegraphics[width=\columnwidth]{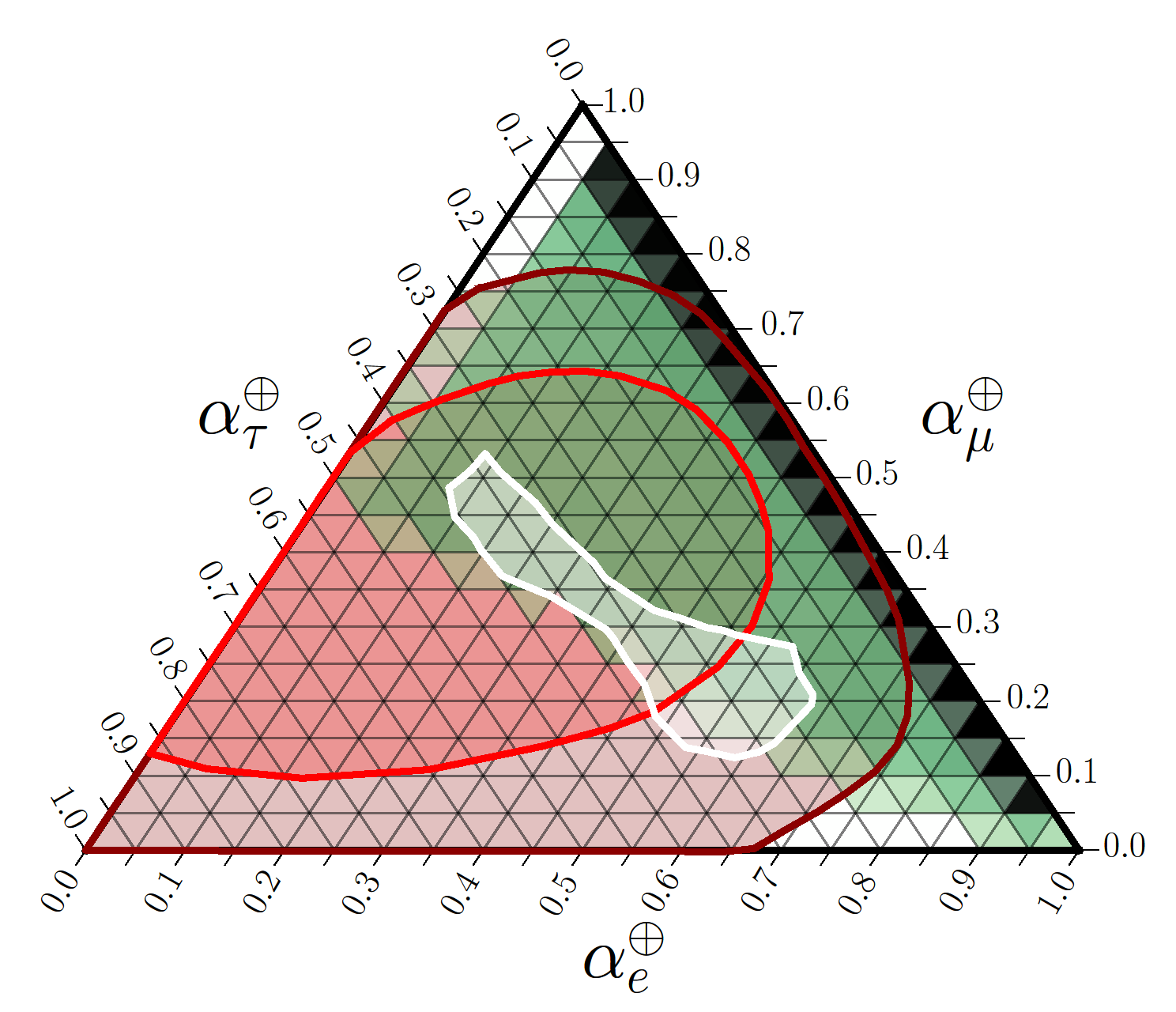}
 \end{center}
\vspace{-2mm}
\caption{The flavor triangle for astrophysical neutrino flavor. Every point in the triangle represents the fraction of astrophysical neutrino flavor. The red enclosed regions are 68\% and 95\% C.L. contours from the IceCube flavor data~\cite{IceCube:2020fpi}. The central white contour represents the predicted flavor ratio on Earth from the standard astrophysical model~\cite{PhysRevLett.115.161303,Bustamante:2015waa}. The green distribution covering the top right half of the figure is the predicted flavor ratio from the neutrino interactions with oscillating ultra-light dark matter field with $g_{\tau\tau}\sqrt{2\rho}/m_\phi=\testlimit$~GeV where the black distribution at the right side of the triangle shows the flavor prediction of the non-oscillating ultra-light dark matter potential.}
\label{fig:fratio}
\end{figure}

Fig.~\ref{fig:fratio} is the computed flavor ratio of astrophysical neutrinos ($\nue:\numu:\nutau$) at detection with nonzero neutrino-dark matter coupling.
Red round contours represent the IceCube data~\cite{IceCube:2020fpi}.
First, we sample the astrophysical neutrino production flavor from the standard astrophysical production mechanisms.
This includes all possible combinations of electron neutrinos and muon neutrinos, or, in symbols, $\nue:\numu:\nutau=x:1-x:0$ with $0<x<1$.
Second, neutrino oscillation parameters are sampled from~\cite{Esteban:2020cvm}.
Finally, the flux with power index $\gamma\sim 2.9$~\cite{IceCube:2020wum} is integrated, and the flavor ratio is computed following~\cite{PhysRevLett.115.161303}. 
The distribution of this standard flavor ratio at detection is shown by a white contour in the center of Fig.~\ref{fig:fratio}.

In the next step, we solve Eq.~\eqref{eq:hamiltonian} assuming nonzero ultra-light dark matter field $\phi(t)$.
First, we start from a non-oscillating field, $\attdiii=g_{\tau\tau}\cdot\sqrt{2\rho}/m_\phi$ by setting $\sin(m_\phi t+\delta)=1$.
This is equivalent to assuming a nonzero constant Lorentz violating field with $\tau-\tau$ sector only.
If the dark matter field amplitude is the IceCube Lorentz violation limit ($=\supbaylimdimiiitattfowo$~GeV), the predicted flavor ratio points are concentrated to the right edge (linear black distribution) which is the outside of the data contour, and it is consistent that such dark matter model is rejected from the data.

Second, we consider the full ultra-light dark matter model, $\attdiii=\phi(t)$. The dark matter field oscillation simulates the smearing effect of astrophysical neutrino flavor content (green distribution covering the top right half), and the flavor ratio prediction is more spread, making some predicted points to be inside of the data contour.
This is because the neutrino - ultra-light dark matter coupling makes an oscillating potential, and it becomes zero with period $(m_\phi t+\delta)/\pi$ in which cases the flavor data cannot be used to look for dark matter.
However, this happens only if the phase is close to $n\pi$ ($n=0, 1, 2, \dots$).

\begin{figure}[t!]
 \begin{center}
 \includegraphics[width=\columnwidth]{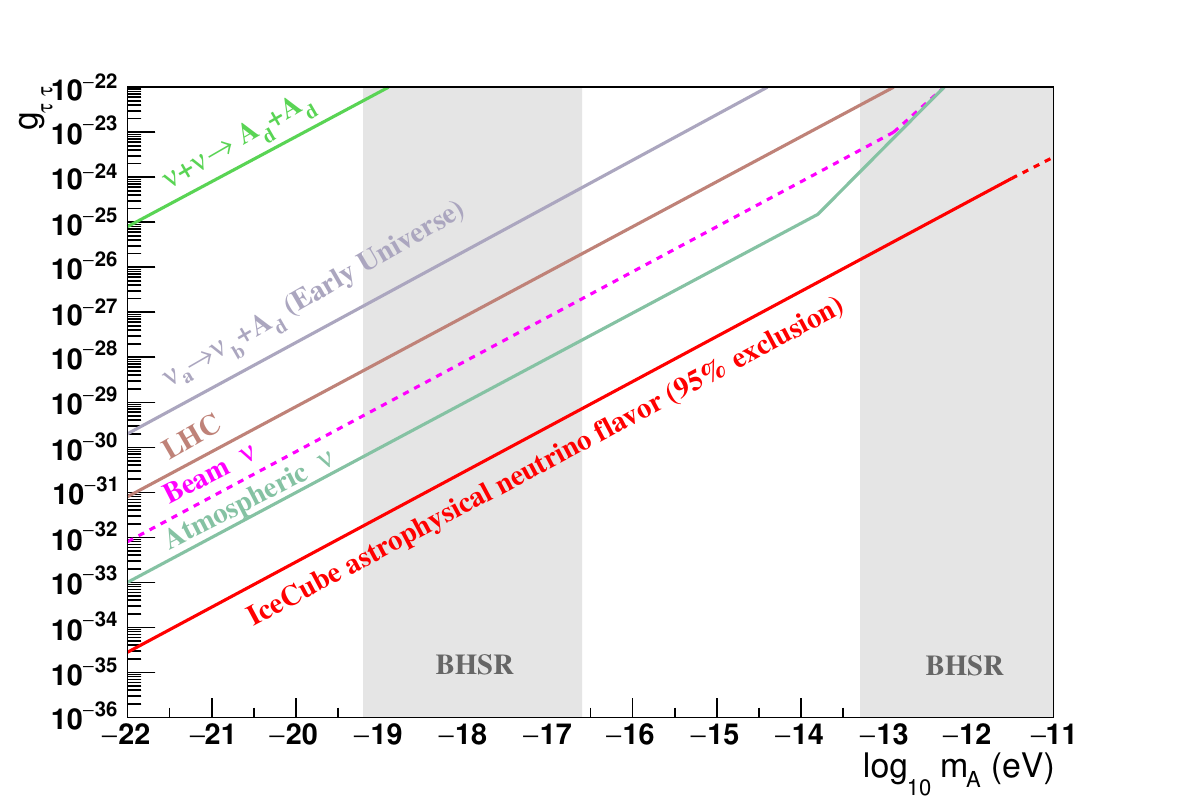}
 \end{center}
\vspace{-2mm}
\caption{Limits and sensitivities of the coupling between astrophysical neutrinos and vector dark matter in $\tau-\tau$ sector. The limits on $g_{\tau\tau}$ is identified with $g'$ from~\cite{Brzeminski:2022rkf} where the plot is adapted. Our limit (red solid line) is overlaid until it starts to violate the adiabaticity condition (red dashed line). Other limits are from early universe (green) $\nu+\nu\rightarrow A_d + A_d$~\cite{Dror:2020fbh,Huang:2017egl}, cosmological neutrino decay (grey)~\cite{Chen:2022idm}, LHC (brown)~\cite{Ekhterachian:2021rkx}, and beam (magenta dashed)~\cite{DUNE:2020fgq} and atmospheric neutrino physics (teal)~\cite{Super-Kamiokande:2014exs,Super-Kamiokande:2014ndf}.}
\label{fig:vector}
\end{figure}

To proceed, we take the following approach:
first, we set the threshold value, and we reject 95\% of predicted flavor ratios from the given ultra-light dark matter amplitude by the 95\% C.L. data contour;
second, the dark matter amplitude is increased until it reaches this threshold value.
By performing this, we derive the limit of the amplitude of oscillating dark matter field is set to $\AmpLimit$~GeV.
Fig.~\ref{fig:vector} is the result. 
Our limit is an order magnitude improvement from current and future neutrino experiments~\cite{Brzeminski:2022rkf,Alonso-Alvarez:2023tii}.
This is mainly due to the higher energy and longer path lengths of astrophysical neutrinos.

\begin{figure}[t!]
 \begin{center}
 \includegraphics[width=\columnwidth]{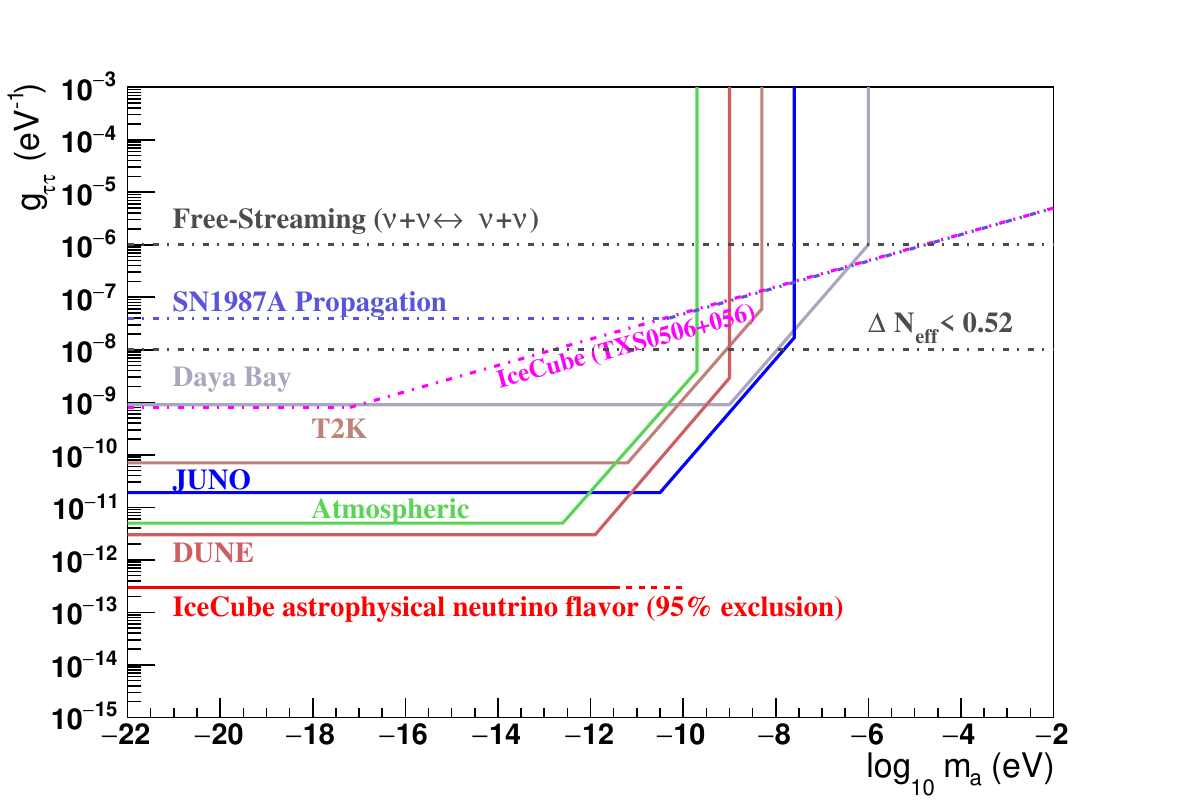}
 \end{center}
\vspace{-2mm}
\caption{
Limits and sensitivities of the coupling between astrophysical neutrinos and axion-like particles in $\tau-\tau$ sector. The limits on $g_{\tau\tau}$ is identified with $g$ from~\cite{Huang:2018cwo} where the plot is adapted, and our limit (red solid line) is overlaid until it starts to violate the adiabaticity condition (red dashed line). Other limits~\cite{Huang:2018cwo} are derived from early universe (grey dash-dotted)~\cite{Huang:2017egl}, SN1987 (blue dash-dotted)~\cite{Kamiokande-II:1987idp,Bionta:1987qt}, TXS0506+056 (magenta dash-dotted)~\cite{IceCube:2018dnn}, Daya Bay (grey)~\cite{DayaBay:2016ggj}, JUNO (blue)~\cite{JUNO:2015zny}, T2K (brown)~\cite{T2K:2018rhz}, DUNE (coral)~\cite{DUNE:2015lol}, and the atmospheric neutrino experiments (green) including KM3NeT~\cite{KM3Net:2016zxf}, Hyper-Kamiokande~\cite{Hyper-Kamiokande:2018ofw}, and IceCube-Gen2~\cite{IceCube-Gen2:2020qha}.
}
\label{fig:axion}
\end{figure}

{\it Axion-like particle dark matter} --- Among all ultra-light dark matter models, axion-like particle dark matter is one of the most popular models~\cite{Marsh:2015xka,Graham:2013gfa,Huang:2018cwo,Alonso-Alvarez:2023tii}.
In this scenario, astrophysical neutrinos can couple with axion dark matter field $a$, via the operator $g_{a\alpha\beta}(\partial_\mu a)\bar\nu_\alpha\gamma^\mu\gamma_5\nu_\beta$.
The field amplitude can be related to the local dark matter density, via $1/2m_a^2a_0^2=\rho$ and $\partial a_0=m_aa_0$, where
\beq
a(t)=g_{a\tau\tau}m_aa_0sin(m_a t+\delta).
\label{eq:axion}
\eeq
The IceCube limit on $\attdiii$ corresponds to $g_{a\tau\tau}\sim\DMA~\eVVmo$. 
This limit can be applied when the axion-like particle mass is small and the axion particles can be treated as a classical field. 
Fig.~\ref{fig:axion} shows the results.
The limit from the IceCube astrophysical neutrino flavor surpasses current and future neutrino experiments~\cite{Huang:2018cwo}. 
Here, mass limit is extended to larger axion mass.
But for the large dark matter mass, oscillations happen very fast and eventually violates the adiabatic condition where the search of galactic axion-like particle stops from astrophysical neutrino flavor.

\begin{table*}
  \begin{tabular}{l|c}
Model & Limits \\
\hline\hline
IceCube Lorentz violation limit & $\attdiii<\supbaylimdimiiitattfowo {\GeVV}$\\
Dark matter potential& $V_{\tau\tau}< \supbaylimdimiiitattfowo{\GeVV}$ \\
Dark matter effective Fermi coupling & $G_F'< \DMFermi$GeV$^{-2}(m_\phi/10^{-20}\eVV)$ \\
Dark matter non-standard interaction & $\epsilon_{\tau\tau}< \DMNSI (m_\phi/10^{-20}\eVV)$ \\
Vector dark matter coupling & $g_{\tau\tau}< \DMV (m_\phi/10^{-20}\eVV)$ \\
Axion dark matter coupling & $ g_{a\tau\tau}< \DMA {\eVVmo}$ \\
  \end{tabular}
\caption{Summary of new physics limits derived from the IceCube astrophysical neutrino flavor.}
\label{tab:1}
\end{table*}

{\it Outlook} --- Table~\ref{tab:1} summarizes all constraints discussed in this Letter.
The Quantum Zeno effect is used to set limits on oscillating ultra-light dark matter field from the astrophysical neutrino flavor.
At this moment, many limits depend on the astrophysical neutrino production model due to limited statistics and precision of particle identification (PID) algorithm.
In the near future, IceCube analyses with larger sample sizes and better particle identification algorithms can remove such dependencies.
For longer timescales, upcoming water neutrino telescopes such as Baikal-GVD~\cite{Baikal-GVD:2018isr}, KM3NeT~\cite{KM3Net:2016zxf} and P-ONE~\cite{P-ONE:2020ljt}, in the mediterranean sea and pacific ocean respectively, will perform additional measurements of astrophysical neutrino flavor~\cite{Song:2020nfh,Liu:2023flr}.
Additionally, dedicated experiments to measure the tau neutrino fraction in the IceCube energy range, such as TAMBO~\cite{Thompson:2023pnl} and Trinity~\cite{Otte:2023osf} are also under development, as well as a high-energy extension of the IceCube array~\cite{IceCube-Gen2:2020qha}

In our analysis we considered the effects of astrophysical neutrinos coupled to ultra-light dark matter~\cite{Navarro:1995iw} through sampling the flavor phase space assuming all possible couplings.
However, when the ultra-light dark matter oscillation length is of the order of the Milky Way radius, it is possible to generate locally high dark matter density.
A precise analysis incorporating such an effect may improve the limit for certain dark matter mass regions~\cite{Karmakar:2020yzn}.
We also neglect parametric resonance effects expected from oscillatory ultra-light dark matter potentials~\cite{Losada:2022uvr} because in most scenarios we do not know the phase of the oscillations.
Finally, here we assume flavor conversion is adiabatic for $m_\phi<\DMAdiabatic$~eV. This condition is conservative, and it can be evaluated more carefully to push the limit to higher dark matter mass regions.

{\it Acknowledgements} --- We thank discussion with Shin'ichiro Ando, Malcolm Fairbairn, Louis Hamaide, Doddy Marsh, and Christoph Terres.
CAA are supported by the Faculty of Arts and Sciences of Harvard University, the National Science Foundation, the Research Corporation for Science Advancement, and the David \& Lucile Packard Foundation.
CAA was also funded by the Alfred P. Sloan Foundation, USA for part of this work.
KF is supported by KAKENHI, Japan.
TK is supported by UKRI STFC, and the Royal Society, UK.

\bibliographystyle{apsrev}
\bibliography{ULDM}

\end{document}